\documentclass{jetpl}
\twocolumn

\lat

\title{Difference of mutant knot invariants and their differential expansion}

\rtitle{Difference of mutant knot invariants and their differential expansion }

\sodtitle{Difference of mutant knot invariants and their differential expansion}

\author{{ L. Bishler$^{a,b,c}$}, { Saswati Dhara$^{d}$}, \ {T. Grigoryev$^{e}$}, \ {A. Mironov$^{a,b,c}$}, \ {A. Morozov$^{b,c,e}$}, \ { An. Morozov$^{b,c,e}$}, \\   { P. Ramadevi$^{d}$}, \ { Vivek  Kumar Singh$^{d}$},\ { A. Sleptsov$^{b,c,e}$}\vspace{1cm}}

\rauthor{{ L. Bishler}, { Saswati Dhara}, \ {T. Grigoryev} et al}

\sodauthor{{ L. Bishler}, { Saswati Dhara}, \ {T. Grigoryev}, \ {A. Mironov}, \ {A. Morozov}, \ { An. Morozov}, \\   { P. Ramadevi}, \ { Vivek  Kumar Singh},\ { A. Sleptsov}}

\address{$^a$ {\small {\it Lebedev Physics Institute, Moscow 119991, Russia}}\\
$^b$ {\small {\it ITEP, Moscow 117218, Russia}}\\
$^c$ {\small {\it Institute for Information Transmission Problems, Moscow 127994, Russia}}\\
$^d$ {\small {\it Department of Physics, Indian Institute of Technology Bombay, Mumbai 400076, India}}\\
$^e$ {\small {\it MIPT, Dolgoprudny, 141701, Russia}}}

\dates{ 07 April 2020}{*}

\abstract{We evaluate the differences of HOMFLY-PT invariants
for pairs of mutant knots colored with representations of $SL(N)$,
which are large enough to distinguish between them.
These mutant pairs include the pretzel mutants,
which require at least the representation,
labelled by the Young diagram $[4,2]$.
We discuss the differential expansion for the differences,
it is non-trivial in the case of mutants, which have the non-zero defect.
The most effective technical tool in this case turns out
to be the standard Reshetikhin-Turaev approach.}

\PACS{ }

\begin{document}

\maketitle

\section{Introduction}

One of the goals of knot theory is to distinguish between different knots.
The most convenient and universal ways to do this is to calculate and compare
some polynomial knot invariants.
Powerful enough while still directly calculable are the HOMFLY-PT polynomials \cite{HOMFLY,PT},
 which are (in a proper normalization) polynomials of two variables $q$ and $A$.
From the physical point of view, they are observables (Wilson loop averages)
in $3d$ Chern-Simons theory.
At the specialization $A=q^N$, these polynomials are observables (Wilson loop averages) in Chern-Simons theory with the gauge group $SU(N)$ or $SL(N)$ and $q:=\exp\Big({2\pi i\over \kappa +N}\Big)$, where $\kappa$ is the coupling constant \cite{Witt}. These polynomials also depend on the representation
$R$ of the gauge group, running along the Wilson loop.

The most difficult knots to distinguish  are the pairs of mutant knots.
These are families of knots which can be transformed into each other
using a special mutation transformation (see section 2).
These knots have the same HOMFLY-PT polynomials in all symmetric
and even all rectangular representations $R$ \cite{mor,mor1} (in fact, these are the representations whose decomposition of the tensor square does not contain non-trivial multiplicities \cite{NRV,Rama1}).
Thus, in order to distinguish these knots, one needs to study mixed representations.
The simplest of them is representation $R=[2,1]$, and it indeed allows one to distinguish between some mutant knots  \cite{NRV,Rama1}.
However, as was explained by H. Morton \cite{mor2} (see also \cite{Rama1}),
there are mutants that possess even higher degree of symmetry.
These mutants are not distinguished (resolved) by the representation $R=[2,1]$
and one needs at least  $R=[4,2]$ to this end.
The study of these mutant knot polynomials and the differences between them is
a challenging problem, interesting both from the point of view of knot theory and
of representation theory.
At the moment no pair of mutants is known, which is {\it not} resolved even by $R=[4,2]$.

The most efficient method to  calculate the HOMFLY-PT polynomials is to apply the Reshetikhin-Turaev (RT) approach, first proposed in \cite{RT}-\cite{RT3} and based on the use of the ${\cal R}$-matrix
for the quantum group $U_q(SL(N))$.
Its topical form, which is sometimes called the modern RT, uses a specific $N$-independent basis for
${\cal R}$-matrices.
It was developed in a series of papers  \cite{GKR1}-\cite{NRZ}, \cite{ModernRT1}-\cite{mut21},
applied to knot polynomials calculations for a variety of knots and links,
and proved to be technically much more powerful in most cases.
However, as we emphasize in this letter, in some problems the original RT formulation turns to be more straightforward and fast,
but the explicit polynomial in variables $q,A$  may not be possible.
This is because the modern RT technique requires knowledge of the Racah matrices, which are very hard to find in the case under consideration.
We provide more details on these approaches in section \ref{secRT}.

We used the RT approach to calculate the differences between the polynomials of mutant knots in representations $R=[3,1]$ and $R=[4,2]$  at some particular values of $N$.
For $[3,1]$, we have managed to do it up to $N=7$.
This allowed us to construct the general answers for any $N$ in this case,
and, hence, to evaluate the corresponding HOMFLY-PT polynomial.
For representation $R=[4,2]$, we  managed to evaluate the differences only for $N=3,\ 4$.
We studied the properties of these differences (see section 4) and their differential expansions.
The differential expansion  \cite{DGR,Diff,Diff1}, \cite{MorK}-\cite{DEnonrect}, which is a relatively new and powerful tool in knot theory, often allows one to guess the unavailable answers for knot polynomials,
study their various properties and generally gives many insights.
It is rather simple for the defect zero knots \cite{MorK} and becomes less trivial in other cases.
Unfortunately, the mutant knots usually have non-zero defects.
Our results obtained here demonstrate that the differential expansion of the mutant knots
exhibits quite interesting properties, we discuss them in section \ref{de}.

\section{Mutant knots }
\label{secmut}

\begin{figure}{}
\centering
\begin{center}
\includegraphics[scale=0.25]{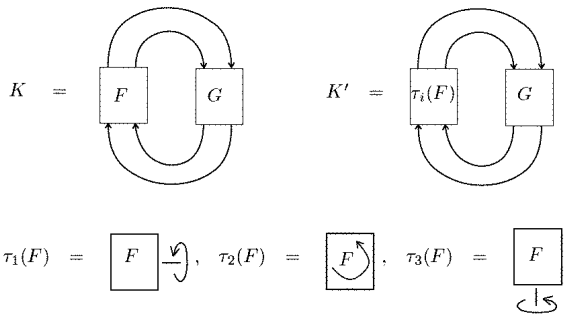}
\end{center}
\caption{1. The mutation procedure\label{fmutation}}
\end{figure}

Let us first discuss in detail what are the mutant knots. Mutant knots are families of knots which are related to each other by a special operation called mutation (see Fig.\ref{fmutation}). This operation means that one cuts a part of a knot 
with two ingoing and two outgoing lines inside
3-sphere. Such a cut portion of the knot, technically referred to as two-tangle, is rotated by 180 deg and glued back.
The resulting knot is a mutant to the initial knot.

Obviously, for these knots to be distinct, both the part in the 3-sphere and outside it should be non-trivial. For this reason, the mutant pairs are knots with many crossings.
The simplest mutant pairs have at least 11 crossings, one of them is formed by the well-known Kinoshita-Terasaka and Conway knots.
However in the case of  11 crossings knots, mutants come only in pairs. We believe that there could be many mutation operations resulting in a family of mutant knots with more than 11 crossings. This is beyond the scope of the present letter.

\begin{figure}{}
\centering
\begin{center}
\includegraphics[scale=0.4]{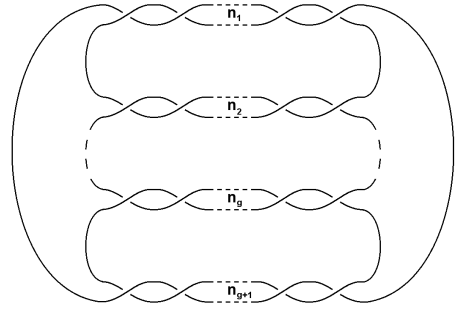}
\end{center}
\caption{2. Pretzel knot $K(n_1,n_2,\ldots,n_g,n_{g+1})$ \label{fpretzel}}
\end{figure}

There is a family of knots called pretzel knots drawn in Fig.2 which includes many new mutants from the mutation operation on any two-tangle.
These are the generalization of the torus knots, which we know a lot about.
The pretzel knots can be put on the genus $g$ surface.
However, unlike torus knots, one puts only two strands on each handle, see Fig.\ref{fpretzel}.
The pretzel knot is parameterized by the numbers of crossings on each handle.

\begin{figure}{}
\centering
\begin{center}
\includegraphics[scale=0.2]{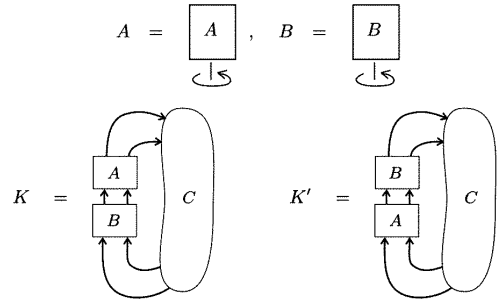}
\end{center}
\caption{3. Mutant knots with higher degree of symmetry from \cite{mor2}\label{fmorton}}
\end{figure}

It is easy to see that interchanging numbers between handles provides exactly a mutation. Thus, starting from genus $4$, the mutant pairs begin to appear. For three and two handles, the mutation gives just the same knot, while for higher genera there are wider sets of mutants.
Note that, among these Pretzel mutant pairs, some get into the class of those possessing even a higher degree of symmetry and distinguishable by representation $R=[4,2]$ only (see Fig.\ref{fmorton} and \cite{Rama1}).

\section{RT $\mathcal{R}$-matrix approach}
\label{secRT}

Reshetikhin-Turaev (RT) approach naturally arises
if the Wilson loop average  \cite{Witt} is evaluated in the temporal gauge \cite{RT4}-\cite{RT5}.
Then each crossing of the  knot diagram (knot projection on the two-dimensional plane)
 is associated  with an $\mathcal{R}$-matrix.
One can start with the universal quantum $\mathcal{R}$-matrix for $U_q(SL(N))$ and calculate the $\mathcal{R}$-matrix in concrete representations using the generators of the quantized universal enveloping algebra:

\begin{equation}
\mathcal{R} = \mathcal{P} q^{\sum_{i,j} a^{-1}_{i,j} h_i \otimes h_j} \overrightarrow{\prod_{\beta \in \Phi^{+}}} {\rm exp}_{q}     \left ( (q - q^{-1}) E_{\beta} \otimes F_{\beta}   \right),
\end{equation}
 where $E_{\beta}$, $F_{\beta}$ and $h_{\beta}$ are the generators of the quantized Universal enveloping algebra:
 \begin{equation}
\begin{array}{ll}
\ [h_i, E_j] = a_{ij} E_j, & [h_i, h_j] = 0,
\\ \
[h_i, F_j] = -a_{ij} F_j, & [E_i, F_j] = \delta_{ij} \frac{q^{h_i}-q^{-h_i}}{q-q^{-1}}.
\end{array}
\end{equation}

Convolution along the edges of $\mathcal{R}$-matrices at the crossings of the
knot diagram and calculating a weighted trace \cite{RT} provides the HOMFLY-PT polynomial.
Unlike the original definition  through the skein relations \cite{HOMFLY},
this approach works equally well for any representation $R$.
However, if one uses the universal $\mathcal{R}$-matrix,
the calculation has to be done separately for each $N$.

A more advanced modern RT approach \cite{ModernRT1}-\cite{ModernRT2}  uses the $\mathcal{R}$-matrix
in the basis of intertwining operators.
When the $\mathcal{R}$-matrix acts
on the tensor product of two representations $V_1$ and $V_2$,
one can consider its decomposition into the irreducible ones, $V_1\otimes V_2=\oplus_Q {\cal M}^Q_{V_1V_2}\cdot Q$.
Since the $\mathcal{R}$-matrix commutes with the co-product \cite{FRT},
its action on the irreducible component is just a number $\pm q^{C_2(Q)}$, \cite{Rev}
where $C_2(Q)$ is the eigenvalue of second Casimir operator in representation $Q$.

Let us now consider an $\mathcal{R}$-matrix in the basis of irreducible representations, or, better to say, in the space of intertwining operators, and realize the knot as a closed $n$-strand braid. Then, the answer for the HOMFLY-PT polynomial $\mathcal{H}^{\mathcal{K}}_V(A,q)$ of the knot ${\cal K}$ colored with representation $V$ is separated into two parts: the one dependent on the group and the one dependent on the knot \cite{ModernRT1}-\cite{ModernRT2}:

\begin{equation}\label{H}
\mathcal{H}^{\mathcal{K}}_V(A,q)=\sum\limits_{Q\in V^{\otimes n}} S^*_Q(A,q) B^{\mathcal{K}}_Q(q).
\end{equation}
From now on, we associate with the representation $Q$ the corresponding Young diagram.
Here $S^*_Q$ is the quantum dimension of the representation $Q$ of the quantum group $U_q(SL(N))$.  The quantum dimension is equal to the Schur polynomial $S_{Q}$ at a special point \cite[sec.7.1.6]{qD}. It does not depend on the knot, and its dependence on $N$ is through the polynomial variable $A=q^N$.
On the other hand, $B^{\mathcal{K}}_Q$  is calculated for a particular knot as a trace of the product of $\mathcal{R}$-matrices and does not depend on $A$ or $N$. What is important, both these factors depend only on the Young diagram $Q$, and all the dependence on $A$ is hidden into the fixed polynomials of $A$, quantum dimensions. This means that this modern RT approach allows one to evaluate the HOMFLY-PT polynomial at all $N$ at once, unlike the standard RT approach that we use in this letter.

Nevertheless, there is a big hindrance in the modern RT approach. It requires, first, a braid representation of the knot which, in the case of mutant knots, requires many strands.
Second, though $\mathcal{R}$-matrices in the space of intertwining operators are simple,
one needs to rotate the basis moving from one to another crossing between different pairs of braids \cite{ModernRT1}-\cite{ModernRT2}. This rotation is provided by the Racah matrices, and evaluating them is very tedious, especially in the case of higher representations \cite{JL1,JL2} and higher number of strands \cite{ModernRT1}, which we are interested in here.

In the calculations reported in this letter,
we do not use the basis of irreducible representations for the $\mathcal{R}$-matrix
and study the knot polynomials
and their differences for particular values of $N$,
which in some cases can be extended to arbitrary $N$.

\section{Mutant knot polynomial differences}
\label{secmd}

In \cite{NRV,Rama1,mut21}, we used the modern RT approach to evaluate mutant differences for the 11 crossing mutant knots. We found that these differences are highly factorizable. They are equal to
\begin{equation}
\Delta H_{[2,1]}^{mutant}=A^\gamma\cdot f(A,q)\cdot {\rm Mt}_{_{[2,1]}}(q)
\end{equation}
where $\gamma$ is an integer, ${\rm Mt}_{_{[2,1]}}(q)$ is a function of only $q$, and
\begin{equation}
f(A,q):=\{q\}^{4}\cdot [3]^2 D_{3}^2D_{2}D_0D_{-2}D_{-3}^2,
\label{ffactor}
\end{equation}
where $[...]$ denotes the $q$-number, $\{q\}:=q-q^{-1}$, and the factors
\begin{equation}
D_k:=Aq^k-A^{-1}q^{-k}
\end{equation}
are called differentials. Note that the $H$ in eqn.(4)  refer to the reduced (normalized) HOMFLY-PT polynomials in variance with the unreduced polynomial (Wilson average) ${\cal H}$ in (\ref{H}).

Now, using the approach described in the previous section and based on the $\mathcal{R}$-matrix at concrete $N$, we evaluated the differences between the polynomials of the mutant knots in representations $[3,1]$ and $[4,2]$, the answers are rather long and can be found in a detailed publication \cite{Mila2} and on a special internet resource \cite{knotebook}.

For representation $[3,1]$, we first managed to calculate the differences for several values $N$. This allowed us to construct the full answers for any $N$ and, hence, to obtain the $[3,1]$-colored HOMFLY-PT polynomial. These differences, however, do not factorize as completely as in the representation $[2,1]$ case. Still there is some structure of factorized differentials:
\begin{equation}
\Delta H_{[3,1]}^{mutant}=\{q\}^4\cdot [4]^2[2]D_{4}D_{3}D_{0}D_{-2}\cdot {\rm Mt}_{_{[3,1]}}(A,q)
\end{equation}
Whenever the differential $D_{-i}$ appears as a factor,
it means the difference vanishes for the $U_q(SL(i))$ group.
Thus we see that
the differences between the mutant knot polynomials disappear for representation $[2,1]$
for the groups $U_q(SL(2))$ and $U_q(SL(3))$,
and for representation $[3,1]$ for the group $U_q(SL(2))$. It is trivial for the group $U_q(SL(2))$, since the difference disappears for any symmetric representation as was explained in the Introduction. It is less trivial for $U_q(SL(3))$ (see \cite{mor}).
The differential $D_i$ with positive $i$
has the same implication for transposed representation $R$.
Representation $R=[2,1]$ does not change under this transposition
and representation $R=[3,1]$ turns into $R=[2,1,1]$.
Thus we see that the differences vanishes for $R=[2,1,1]$ for the group $U_q(SL(4))$.

For the representation $[4,2]$, we were not able to construct the universal answer for all $N$. We, however, managed to calculate the answers for pretzel mutant knots in the case of $U_q(SL(3))$ and $U_q(SL(4))$ groups.

\section{Differential expansion}
\label{de}

The colored HOMFLY-PT polynomial possesses an additional structure called differential expansion (DE) \cite{DGR,Diff,Diff1}, \cite{MorK}-\cite{DEnonrect}, which is related with the representation theory \cite{Diff1,Diff2}.
The simplest example of DE   appears already in the fundamental representation:
since for the abelian $U(1)$ Chern-Simons theory, i.e. for $A=q$, the reduced
polynomial in the topological framing is trivial, we have
\begin{equation}
H_{[1]}(A,q)= 1+D_1D_{-1}\cdot F_{[1]}(A,q)
\end{equation}
with a new, simpler, Laurent polynomial $F_{[1]}(A,q)$.
Continuing further and  looking at other $N$, one comes to the general structure of expansion
of colored polynomials in products
of the knot-independent combinations $Z_R^Q$ of various differentials $D_k$:
\begin{equation}
H_{R}^K(A,q)= \sum_{Q\in M_R} Z_R^Q(A,q) \cdot F_{Q}^K(A,q)
\end{equation}

Important parameter for the differential expansion is the defect $\delta_K$ of a knot $K$.
It is defined by degree of the Alexander polynomial,
i.e. the specialization of the fundamental HOMFLY-PT polynomial at $A=1$:
\begin{equation}
H_{[1]}^K(A,q)\Big|_{A=1}=\sum_{j=-\delta_K-1}^{\delta_k+1}a_jq^{2j}
\end{equation}
The differential expansion is more involved in the case of non-vanishing defect, and this is exactly the case for mutant knots. A general theory of differential expansion in this case will be reported elsewhere, here we just discuss a concrete problem arising for the mutant pairs of knots.

Let us consider the difference of differential
expansions of the HOMFLY-PT polynomials in the
mutant pair.  We denote this difference by $\Delta$. One can assume that contributing to it are only the pairs of non-diagonal composite representations, like $X_2:=([2],[1,1])\oplus ([1,1],[2])$ and $X_3:=([3],[2,1]) \oplus ([2,1],[3])$. We, however, allow also an additional adjustment of some DE coefficients, which remains unobservable in the leading order, but can show up for higher representations, this adjustment will be denoted by small $\delta$.  In the first mixed representation, from \cite[eq.(106)]{Rama1} and \cite[eqs.(14)-(17)]{DEnonrect}, one gets an expression, depending on two unknown functions of the form
$$
\Delta H_{[2,1]} =
\frac{[3]}{[2]^2}\Big(\underbrace{D_0^2}_{{\footnotesize \times} 0}
+ \underbrace{[3]D_2D_{-2}}_{{\footnotesize \times}\{q\}^4[2]^2\cdot\delta F_{[1]}}\Big)
\oplus \underbrace{\{q\}^4[3]^2D_2D_{-2}}_{{\footnotesize \times}(\Delta F_{X_2} - \delta F_{[1]})}
=
$$
\begin{equation}
=\{q\}^4\cdot [3]^2D_3^2D_2D_0D_{-2}D_{-3}^2 \cdot {\rm Mt}_{_{[2,1]}}
\end{equation}

\noindent
$\delta F_1$ denotes a possible redistribution of the coefficients
between the different terms of the differential expansion for two mutants,
which does not affect the r.h.s., but can show up in the mutant difference
for higher representations.
It is natural to assume that it vanishes, but we keep this option open.

Similarly, for the next mixed representation
{\footnotesize
$$
\Delta H_{[3,1]} =
\frac{[4]}{[3]}\Big(\underbrace{D_1D_0}_0
+ \underbrace{\frac{[4]}{[2]}D_3D_{-2}}_{{\footnotesize \times}[2]^2\{q\}^4\cdot\delta F_{[1]}}\Big) \oplus
\underbrace{\{q\}^4[4]^2[2]D_3D_{-2}}_{{\footnotesize \times}(\Delta F_{X_2} - \delta F_{[1]})}
+
$$
$$
+ \frac{[4]}{[3]^2}\Big(\!\underbrace{D_3D_1^{\overline 2}D_0}_{{\footnotesize \times} 0}
+ \underbrace{[4][2]D_4D_3\overline{D_0}D_{-2}}_{{\footnotesize \times}\{q\}^4[3]^2\cdot\delta F_{[2]}}\Big)
\oplus
\underbrace{\{q\}^4 [4]^2[2]D_4D_3\overline{D_0}D_{-2}}_{
{\footnotesize \times}(\Delta F_{X_3} - \delta F_{[2]})}
$$
}
\begin{equation}
=\{q\}^4\cdot [4]^2[2]D_4D_3D_0D_{-2}  \cdot {\rm Mt}_{_{[3,1]}}
\end{equation}
Now we have a problem:
everything in the last line is divisible by $D_4$,
but $\Delta F_{X_2}\neq 0$ is not.

There are at least two possible ways out.
One possibility is to allow  $\delta F_{[1]}\neq 0$.
For example, take
$$
\Delta F_{X_2} = D_3^2D_0D_{-3}^2 \cdot {\rm Mt}_{_{[2,1]}} \\
\delta F_{[1]} = D_3D_2D_1D_{-3}^2 \cdot {\rm Mt}_{_{[2,1]}}
$$

\noindent
so that $\Delta F_{X_2}-\delta F_{[1]}=-[2]\{q\}^2D_3D_{-3}^2\cdot {\rm Mt}_{_{[2,1]}}$.
Then we get:
$$
\{q\}^4[4]^2[2]D_2D_{-2}\Big(\Delta F_{X_2}
-\underbrace{\Big(1-\frac{1}{[3]}\Big)}_{\frac{[4]}{[3][2]}} \delta F_{[1]}\Big)=
$$
\vspace{-0.3cm}
$$
= \{q\}^4[4]^2[2] D_2D_{-2} D_3D_{-3}^2
\underbrace{\left(D_3D_0 -\frac{[4]}{[3][2]}D_2D_1\right)}_{\frac{D_4D_{-1}}{[3]}}\cdot {\rm Mt}_{_{[2,1]}}
$$

\noindent
which is divisible by $D_{4}$.
However, now arises a new potential problem: \ we get $D_4D_{-1}$ rather than $D_4D_0$,
but overlined $D_0$ can actually be absent from the differential expansion for non-vanishing defect.
In any case, still many more things need to match...

Another possibility is to note that $F_{[3,1]}$,
which was not taken into account in above differences,
can also be different for the two mutants.
In the case of $H_{[2,1]}$, we had $\Delta F_{[2,1]}=0$,
because it also enters the expansion of rectangular $H_{[2,2]}$
which does {\it not} distinguish mutants, i.e. $\Delta H_{[2,2]}=0$.
However, $H_{[3,3]}$
contains contributions from {\it two}
non-rectangular structures $F_{[3,1]}$ and $F_{[3,2]}$,
thus $\Delta F_{[3,1]}$ and $\Delta F_{[3,2]}$ can be non-vanishing
and compensate each other in the vanishing rectangular $\Delta H_{[3,3]}$.

\section{Conclusion}

This letter is a brief summary of our results for
the HOMFLY-PT polynomials of the mutant knots.
These polynomials and especially the differences between them are of great interest
from many points of view.
We  managed to construct the differences between these polynomials
in representation $R=[3,1]$ for all 11-crossing mutant knots.
These are much less structured than in the representation $[2,1]$ case,
nevertheless the representation dependence is not quite  trivial.
We also studied the differential expansion of these differences, which is related to their representation properties.
In particular, we realize a subtle point in the differential expansion of mutants that requires further development in the case of knots with non-vanishing defect. It remains to be seen what happens for higher representations.

We also evaluated differences between the polynomials of mutant knots in representation $[4,2]$, but only for the $U_q(SL(3))$ and $U_q(SL(4))$ groups, which did not allow us to find the general answer, though allowed us to distinguish between mutants. To find the whole HOMFLY-PT invariant in these cases, one needs either some new approaches  or
serious optimization of computer programs.

\section*{Acknowledgements}

Our work is supported in part by the grant of the
Foundation for the Advancement of Theoretical Physics ``BASIS" (L.B., A.M.'s, A.S.),
by President of Russian Federation grant MK-2038.2019 (L.B., An.M.),
by  RFBR grants 19-01-00680 (A.Mir.), 19-02-00815 (A.Mor.), 20-01-00644 (An. Mor., A.S.)
18-31-20046-mol-a-ved (A.S.),
by joint grants 19-51-50008-YaF-a (L.B., A.Mir., An.Mor.), 19-51-53014-GFEN-a, 18-51-05015-Arm-a, 18-51-45010-IND-a (L.B., A.M.'s, A.S.), PR, VKS and SD acknowledge DST-RFBR grant (INT/RUS/RFBR/P-231) for support.
The work was also partly funded by RFBR and NSFB according
to the research project 19-51-18006 (A.Mir., A.Mor., An.Mor.). VKS would like to thank IISER, Pune (India)  where part of
this work was done during his visit as visiting fellow.  A.Mir., A.Mor. and PR also acknowledge
the hospitality of KITP and partial support by the National Science Foundation under Grant No. NSF PHY1748958.

\end{document}